\documentclass[aps,twocolumn,groupedaddress,showpacs]{revtex4}

\newcommand{\E}{{\cal E}}

\begin{document}

\title{Quantum authentication of classical messages}


\author{Marcos Curty}
\email[]{mcurty@com.uvigo.es}

\author{David J. Santos}
\email[]{dsantos@com.uvigo.es}
\affiliation{ETSIT, Universidad de Vigo, Campus Universitario s/n,
E-36200 Vigo (Spain)}

\date{\today}

\begin{abstract}
Although key distribution is arguably the most studied context on
which to apply quantum cryptographic techniques, message
authentication, i.e., certifying the identity of the message
originator and the integrity of the message sent, can also benefit
from the use of quantum resources. Classically, message
authentication can be performed by techniques based on hash
functions. However, the security of the resulting protocols
depends on the selection of appropriate hash functions, and on the
use of long authentication keys. In this paper we propose a
quantum authentication procedure that, making use of just one
qubit as the authentication key, allows the authentication of
binary classical messages in a secure manner.  \end{abstract}


\maketitle

\section{Introduction}

As computer networks spread worldwide with users accessing them via
millions of different terminals, information protection becomes more and
more relevant. This challenge of providing adequate information protection
is closely related to the basic tasks of cryptography: authentication
and secrecy \cite{DENNING_1983,SCHNEIER_1996}. During the last decade it
has been shown that information has a physical, not only mathematical,
dimension and, as such, can be studied making use of Quantum Theory. This
has given birth to the research field known as Quantum Information Theory
(QIT) (see, e.g., \cite{GRUSKA_1999,LO_1999_a,NIELSEN_2000}). Quantum
Cryptography (QC), first introduced by Wiesner \cite{WIESNER_1983} and Bennett
and co-workers \cite{BENNETT_1982}, is, with Quantum Computation, one of the
most remarkable applications of QIT.  The information security provided
by QC is based on fundamental properties of Quantum
Mechanics, instead of on unproven assumptions concerning the computational
complexity of some algorithms (as it is the case of most of Classical
Cryptography), and therefore brings a whole new dimension to security
in communications.  Over the last few years there have been several
experimental demonstrations of the feasibility of QC
\cite{BENNETT_1992,MULLER_1993,MARAND_1995,HUGHES_1995,BUTTLER_2000,TITTEL_2000,NAIK_2000,JENNEWEIN_2000}
which seem to indicate that the prospects for its future mainstream use
are good.

QC involves several topics, and although Quantum Key Distribution
(QKD) \cite{BENNETT_1984,EKERT_1991,BENNETT_1992_a} is arguably
the most studied one, the necessity to combine QKD protocols with
classical authentication methods has motivated recent
investigations on the achievement of key verification
\cite{LJUNGGREN_2000,ZENG_2000} and user authentication
\cite{CREPEAU_1995,DUSEK_1999,barnum_99,ZENG_2000_a,JENSEN_2000,zhang_00}
in a quantum-mechanical secure manner. Key verification consists
of assuring that the parties of a key-distribution scheme are the
legitimate ones, and that the key established is authentic.  User
authentication (also called user identification) allows a
communicator to prove his/her identity, often as the first step to
log into a system.  One potential insecurity of user
authentication consists of assuming that once the log-in process
has concluded, the transmission remains authentic for the rest of
the communication. This assumption strongly depends on the level
of security provided by the channel used.  Classical Cryptography
solves this weakness employing message authentication codes
(MACs), which enable parties owning a shared secret key to achieve
data integrity. A MAC, also known as a data authentication code,
is essentially a scheme specified by two algorithms: an encoding
or tagging algorithm (possibly stochastic), and a decoding or
verification algorithm. When the sender (Alice) wants to send a
certified message to a recipient (Bob), she computes, employing
the encoding algorithm, a tag (as a function of the message and a
secret key previously shared) and appends it to the message. On the
reception side, Bob verifies the authenticity of the tag by means
of the specified decoding procedure, which depends on the message,
tag, and shared key.  This algorithm returns a bit indicating
when Bob must regard the message as authentic and accept it as
coming from Alice, and when he must discard it.  Wegman and Carter
\cite{CARTER_1979,WEGMAN_1981} described a message-authentication
scheme whose security is information-theoretic, rather than based
on computational assumptions. Their technique uses a hash function,
selected from a Universal Hash Family, to compress the message to be
certified into a smaller string of bits. Then this string is encrypted
to produce the tag.

Recently, Barnum \cite{barnum_2001} has addressed the problem of
authenticating quantum messages. In his protocols the
authentication key is used to select a quantum error-detection
code (QEDC) from a given set. A quantum state is encoded in one of
these codes, and the state is rejected as inauthentic if an error
is detected by the recipient. The geometry of the set of QEDCs is
chosen such that it ensures that the probability of undetected
forgery is less than the classical bound (inverse of the square
root of the number of keys).

In this paper we study how the use of quantum resources can
improve the authentication of classical messages. Specifically, we
present a broad class of quantum authentication schemes that,
unlike classical MACs, which need at least two secret bits to
achieve a probability of forgery less than one, provide secure
data integrity when only one-qubit key is shared between the
communicating partners.

The paper is organized as follows. In Section~\ref{PROTOCOL} we
describe a class of quantum message-authentication codes that
requires just one qubit as the key to authenticate binary
messages. In Section~\ref{SECURITY} we analyze the security of
these protocols against various attacks of increasing severity.
First, we analyze the no-message attack, in which the sender has
not initiated the transmission (there is no message in the
channel), and the adversary (Eve) attempts to prepare a message
with the goal of passing Bob's verification test. Then, we analyze
more subtle attacks, those in which Eve has access to what is
transmitted.  We also discuss, in Section~\ref{SECURITY}, how the
security of the protocol is modified if the authentication keys
are reused. Finally, we present our conclusions in
Section~\ref{CONCLUSIONES}.

\section{Quantum message-authentication codes}
\label{PROTOCOL}

Suppose Alice needs to send a certified classical message to Bob.
The goal is to make Bob confident about the authenticity of the message
and sender. The protocols described in this section require a quantum
channel, so the first task consists of assigning a quantum state to
each possible classical message. This decision needs no secrecy and
can be made openly. We will discuss the simple case of binary messages
(one-bit long).  Thus, there are only two possible messages, `0' and
`1', to which we assign the pure quantum states $|\phi_{0}\rangle$ and
$|\phi_{1}\rangle$, respectively.  In order to guarantee Bob's perfect
extraction of information from these states and to make authentication
possible, they cannot be selected arbitrarily, but must be orthogonal,
$\langle\phi_{i}|\phi_{j}\rangle=\delta_{ij}$, with $i,j\in\{0,1\}$;
and must contain, as in any authentication method, some tag information
to be checked by Bob. We will assume that they belong to a two-qubit
state space (a four-dimensional Hilbert space) ${\cal E}$. This can be
seen as if the first qubit carried the message information,
and the second qubit carried the tag. As for the secret authentication
key, we will assume that Alice and Bob share a two-qubit maximally
entangled state: Each owns one qubit of a publicly-known singlet state
$|\psi\rangle_{AB}=\frac{1}{\sqrt{2}}(|01\rangle_{AB}-|10\rangle_{AB})$.

The authentication procedure goes as follows: When Alice wants to send a
certified bit $i$, she prepares two qubits in the state $|\phi_{i}\rangle$
and performs the following encoding operation on her part of
$|\psi\rangle_{AB}$ and on the message:

\begin{equation}
E_{A{\cal E}}=|0\rangle\langle0|_A 1_{\cal E}+|1\rangle\langle 1|_A U_{\cal E},
\label{ENCODING}
\end{equation}

\noindent where $U_{\cal E}$ is some publicly-known unitary quantum
operation. Basically, the result of this encoding operation can be
seen as performing  (second term in (\ref{ENCODING})) or not (first
term in (\ref{ENCODING})), depending on the state of Alice's qubit of
the shared key, a unitary operation, $U_{\cal E}$, on the quantum state
$|\phi_i\rangle$. This could also be accomplished with a previously
shared classical bit acting as a key. The singlet can be seen as a
superposition of all possible classical key states.

After performing this tagging operation, the state of the global
system (Alice+Bob+message) is

\begin{equation}
\frac{1}{\sqrt{2}}\left(
|01\rangle_{AB}|\phi_i\rangle- |10\rangle_{AB}U_{\cal E}|\phi_i\rangle
\right).
\label{DAVID_1}
\end{equation}

\noindent Using the density operator formalism, the state of the
authenticated message that Alice sends to Bob can be obtained from (\ref{DAVID_1})
performing the partial trace over the Alice+Bob variables. In density
operator terms, this state is given by

\begin{equation}\label{prot1}
\rho^\prime=\frac{1}{2}(\rho_{i}+U_{\cal E}\rho_{i}U_{\cal E}^{\dag}),
\end{equation}

\noindent where $\rho_{i}=|\phi_{i}\rangle\langle\phi_{i}|$.  On the
reception side, Bob decodes the information sent by Alice performing
the decoding operation

\begin{equation}
D_{B{\cal E}}
=|0\rangle\langle0|_B U_{\cal E}^\dagger+|1\rangle\langle 1|_B 1_{\cal E}
\end{equation}

\noindent on his part of $|\psi\rangle_{AB}$ and the message
received. Finally, Bob performs an orthogonal measurement on the
space ${\cal E}$. Since this space is four-dimensional, and we
have imposed the states $|\phi_0\rangle$ and $|\phi_1\rangle$ to
be orthonormal, we can perform this measurement on the orthonormal
set $\{|\phi_i\rangle;\, i=0,\ldots,3\}$, where $|\phi_2\rangle$
and $|\phi_3\rangle$ are two extra orthonormal states.  If the
result of such a measurement is one of the two first elements of
the set, Bob should assume that no forgery has taken place, and
therefore obtain the classical message sent to him. If this is not
the case, he rejects the message received.

\section{Security analysis}
\label{SECURITY}

The class of quantum protocols of the previous section provides perfect
deterministic decoding, i.e., the quantum key $|\psi\rangle_{AB}$ and
the quantum ciphertext $\rho^\prime$ uniquely determine the classical
message sent, $\rho_i$. This means that these protocols would fail only
if Bob accepted a message as an authenticated one when that is not the
case (due to the unnoticed action of Eve). When dealing with forgery
strategies we must consider two main types of attacks: The no-message
attack, and the message attack. The first one is the simplest: Before
Alice's sending any message to Bob, Eve attempts to prepare a quantum
state that passes the decoding algorithm. The message attack is more
subtle and severe: Eve could access authentic messages transmitted, and
try to produce a forged message based on the information gained. The
purpose of this section is to analyze both families of attacks, and
obtain the class of unitary operations $U_{\cal E}$ that makes them
unsuccessful. In the following discussion we will consider the ideal
scenario of an error-free quantum channel.

\subsection{No-message attack}\label{attack1}

Suppose Eve prepares a normalized pure quantum state
$|\epsilon\rangle\in{\cal E}$ and sends it to Bob trying to
impersonate Alice. In the most general case, this inauthentic pure
quantum message can be described as
$|\epsilon\rangle=\sum^{3}_{i=0}e_{i}|\phi_{i}\rangle$. When Bob
receives this quantum message he cannot know that it comes from a
forger, so he follows the procedure explained in the previous
section: He performs a decoding operation and then an orthogonal
measurement over the set $\{|\phi_{i}\rangle;\, i=0,\ldots,3\}$.
Before this measurement takes place, the state of the message can
be described by $\rho_{E}^\prime=(\rho_{E}+U_{\cal
E}^{\dag}\rho_{E}U_{\cal E})/2$, where
$\rho_{E}=|\epsilon\rangle\langle\epsilon|$. As we have seen, Bob
rejects the message if the result of his measurement is one of the
last two elements of this basis; therefore, the probability $P_f$
that Eve deceives Bob is:

\begin{equation}\label{gen}
P_f=\sum_{i=0}^{1}
\langle\phi_{i}|\rho_{E}^\prime|\phi_{i}\rangle=
\frac{1}{2}\sum_{i=0}^{1}
\left(|e_{i}|^{2}+
|\langle\epsilon|U_{\cal E}|\phi_{i}\rangle|^{2}\right).
\end{equation}

\noindent This quantity depends both on Eve's strategy and on the quantum
operation $U_{\cal E}$. The normalization of $|\epsilon\rangle$ and
the unitarity of $U_{\cal E}$ make both terms on the right side of
(\ref{gen}) to be less or equal than 0.5.  The first term depends entirely
on Eve's decision, and, to be 0.5, $e_{2}$ and $e_{3}$ must be zero. We
will assume that Eve selects $|\epsilon\rangle$ such as this
condition is fulfilled. Let
us focus on the second term, $1/2\sum_{i=0}^1|\langle \epsilon|U_{\cal
E}|\phi_i\rangle|^2$. First, let us write the matrix representation of
$U_{\cal E}$ in the block form

\begin{equation}
U_{\cal E}=
\left(
\begin{array}{cc}
M_{0} & M_{1} \\
M_{2} & M_{3} \\
\end{array}
\right),
\end{equation}


\noindent where the $M_i$ are $2\times 2$ complex matrices. With this
notation, the second term in the right side of (\ref{gen}) can be
written as

\begin{eqnarray}\label{gen2}
\frac{1}{2}
& &\left[
\left(|\bar{M}_{0}^{0}|^{2}-
|\bar{M}_{0}^{1}|^{2}\right)|e_{0}|^{2}+\right.\nonumber \\
& &\left.2|\bar{M}_{0}^{1}\bar{M}_{0}^{0\dag}||e_{0}|
\sqrt{1-|e_{0}|^{2}}\cos{\theta_{E}}
+|\bar{M}_{0}^{1}|^{2}
\right],
\end{eqnarray}

\noindent where $\bar{M}_{i}^{j}$ represents the $j$-row of the
$i$-block of $U_{\cal E}$, and $\theta_{E}$ is an angle that depends entirely on
Eve's choice of her state. Eve's goal is to make $P_f$ as big as
possible, so the worst case for Alice and Bob occurs when Eve chooses
$\theta_E=2\pi{}k$ with $k$ any integer, and a $|e_0|$ that maximizes
(\ref{gen2}) for a given $U_{\cal E}$. We can distinguish between two cases:

\begin{enumerate}
\item If $|\bar{M}_{0}^{1}\bar{M}_{0}^{0\dag}|=0$, the maximum of
(\ref{gen2}) is strictly less than 0.5 when
$|\bar{M}_{0}^{0}|^{2}<1$ and $|\bar{M}_{0}^{1}|^{2}<1$.

\item If $|\bar{M}_{0}^{1}\bar{M}_{0}^{0\dag}|\neq 0$, the maximum of
(\ref{gen2}) is strictly less than 0.5 when

\begin{equation}
\frac{1}{2}x{}\Bigg\{1+\bigg(\frac{x}{y}\bigg)\bigg[1+\bigg(\frac{x}{y}
\bigg)^{2}\bigg]^{-1/2}\Bigg\}+ \frac{1}{2}y\bigg[1+\bigg(\frac{x}
{y}\bigg)^{2}\bigg]^{1/2}+z<1,
\label{EC_GORDA}
\end{equation}

\noindent where the real variables $x,y$ and $z$
are $|\bar{M}_{0}^{0}|^{2}-|\bar{M}_{0}^{1}|^{2}$,
$2|\bar{M}_{0}^{1}\bar{M}_{0}^{0\dag}|$ and $|\bar{M}_{0}^{1}|^{2}$,
respectively.

\end{enumerate}

\noindent Note that, in both cases, Alice and Bob can select $U_{\cal E}$ such
that its $M_0$ block makes $P_f<1$ independently of Eve's choice of
$|\epsilon\rangle$.

Finally, in this subsection we have assumed that Eve prepares a
pure state $|\epsilon\rangle$; however, she could have prepared a
general mixed state $\rho_E=\sum_{i=0}^3 p_i
|\epsilon_i\rangle\langle \epsilon_i|$, with $\sum_{i=0}^3p_i=1$. From
what we have shown in this subsection, it is straightforward to
see that if $U_{\cal E}$ is selected satisfying the conditions above,
then also in this case $P_f<1$. In fact, we can further show that,
with the appropriate selection of $U_\E$, $P_f$ can be made at
most $1/2$. According to (\ref{gen}), $P_f$ can be written as
$P_f=\mbox{Tr}\left({\rho_E^\prime} P\right)$, where
${\rho_E^\prime}=\left(\rho_E+U_\E^\dagger\rho_E U_\E\right)/2$ and
$P=|\phi_0\rangle\langle \phi_0|+|\phi_1\rangle\langle
\phi_1|$. Using the properties of the trace operator,

\begin{equation}
P_f=\mbox{Tr}\left(\rho_E Q\right)/2,
\end{equation}

\noindent where $Q=U_\E P U_\E^\dagger+P$ is a positive operator known
to Eve, and with maximum eigenvalue $\lambda_{max}\geq 1$. Therefore,
the maximizing $\rho_E$ is any eigenvector corresponding to
$\lambda_{max}$, and thus $P_f=\lambda_{max}/2$.  Finally, it is easy
to see (see, e.g., \cite{HORN_1985}) that choosing $U_\E$ such that it
takes $P$ to its orthogonal complement makes $\lambda_{max}=1$, and
therefore, as predicted, $P_f=1/2$.

\subsection{Message attack}

As we have seen, this is a more subtle and severe class
of attacks. Instead of directly forge a quantum message and send
it to Bob, Eve could wait for Alice's original messages and try to
manipulate them. Thus, Eve's goal is to convert authentic messages into
others passing Bob's test. In the simple case we are dealing with
(binary messages), this implies converting $|\phi_{0}\rangle$ into
$|\phi_{1}\rangle$ and vice versa.

In order to simplify the analysis, and without loss of generality,
we will distinguish between two types of message attacks.  In the
first one, Eve, based on the knowledge of all the public aspects
of the quantum authentication scheme used, determines a quantum
operation and applies it to any message sent by Alice. This
quantum operation can be described by a trace-preserving
completely-positive (TPCP) map. In the second class of attacks,
Eve also tries to extract information, by means of the appropriate
measurement of the message in the channel, that allows her to
prepare a different message that Bob regards as authentic.

\subsubsection{TPCP map}

Consider that  Alice sends to Bob a quantum message $|\phi_{i}\rangle$,
with $i\in\{0,1\}$, and Eve performs an arbitrary TPCP map, ${\cal
M}$, on it. The new state in the channel is $\rho_{E}^\prime={\cal
M}(\rho^\prime)$, with $\rho^\prime$ given by (\ref{prot1}). Eve chooses
${\cal M}$ such that the decoding procedure performed by Bob on the
resulting state lead to the state $|\phi_j\rangle$, with $j\in \{0,1\}$,
and $j\neq i$. Owing to the pure character of the states $|\phi_0\rangle$
and $|\phi_1\rangle$, this can only be done with certainty if ${\cal
M}$ is a unitary operation, that we will write as $U_E$. For this
kind of operation, the probability, $P_f^\prime(i)$, of Eve achieving her
goal is $\langle\phi_j|\rho_E^{\prime\prime}|\phi_j\rangle$, where
$\rho_E^{\prime\prime}$, Bob's decoded state, is

\begin{equation}
\rho_E^{\prime\prime}=\frac{1}{2}
\left(
U_E \rho_i U_E^\dagger + U_{\cal E}^\dagger U_E U_{\cal E} \rho_i
U_{\cal E}^\dagger U_E^\dagger U_{\cal E}
\right),
\end{equation}

\noindent with $\rho_i=|\phi_i\rangle\langle\phi_i|$. Therefore,

\begin{equation}\label{man1}
P_f^\prime(i)= \frac{1}{2}(|\langle\phi_{j}|U_E|\phi_{i}\rangle|^{2} +
|\langle\phi_{j}|U_{\cal E}^{\dag}U_{E}U_{\cal E}|\phi_{i}\rangle|^{2}).
\end{equation}

\noindent If Alice prepares the state $|\phi_i\rangle$ with probability
$p_i$, the overall probability of, employing a TPCP, substituting
an authentic message with a different one that passes Bob's test is
$P_f^\prime=\sum_i p_i P_f^\prime(i)$.  This probability is one if $U_{E}$
simultaneously satisfies, up to some arbitrary global phase factors,
the following two pairs of conditions:

\begin{equation}
|\phi_{j}\rangle=U_{E}|\phi_{i}\rangle,
\label{PAR_1}
\end{equation}

\noindent and

\begin{equation}
|\phi_{j}\rangle=U_{\cal E}^{\dag}U_{E}U_{\cal E}|\phi_{i}\rangle,
\label{PAR_2}
\end{equation}

\noindent $\forall$ $i,j\in\{0,1\}$, with $i\neq{}j$.  The orthogonality
between $|\phi_0\rangle$ and $|\phi_1\rangle$ allows Eve to always fulfill
one of the two pairs of conditions independently of the particular $U_{\cal E}$
employed by Alice and Bob. Let us assume that Eve
selects $U_{E}$ such that (\ref{PAR_1}) is satisfied.  This selection
makes $U_{E}$ to have, in the orthonormal base
$\{|\phi_{i}\rangle; i=0,\ldots,3\}$, the following  block
representation:

\begin{equation}\label{man2}
U_E=\left(
\begin{array}{cc}
M_{0}^{E} & 0 \\
0 & M_{1}^{E} \\
\end{array}
\right)
\end{equation}


\noindent with $M_{0}^{E}=e^{i\alpha}S(\beta)\sigma_x$, where
$\alpha$ is an arbitrary phase, $\sigma_x$ is the standard Pauli
matrix, and $S(\beta)$ is a phase-shift operation, whose matrix
representation is

\begin{equation}
S(\beta)=\left(
\begin{array}{cc}
1 & 0\\
0 & e^{i\beta}
\end{array}
\right);
\end{equation}

\noindent and $M_1^E$ is any $2\times 2$ unitary matrix. Now, if
we further demand the fulfillment of (\ref{PAR_2}), the matrix
elements of $U_{\cal E}$ and $U_{E}$ must obey
$\langle\phi_{k}|U_{\cal E}|\phi_{i}\rangle=\sum_{l=0}^{3}
\langle\phi_{k}|U_{E}|\phi_l\rangle\langle\phi_{l}|U_{\cal
E}|\phi_{j}\rangle$ $\forall{}k\in\{0,\ldots{},3\}$. With the
notation of $U_{E}$ introduced in equation (\ref{man2}), this
implies that $M_{0}^{0},M_{0}^{1},M_{2}^{0}$ and $M_{2}^{1}$,
where $M_{i}^{j}$ represents the $j$-column of the $i$-block of
$U_{\cal E}$, must satisfy
$M_{0}^{0}=e^{i\gamma}S(\delta)\sigma_xM_{0}^{1}$ and,
$M_{2}^{0\dag}M_{2}^{1}=0$ or $M_{2}^{0}=e^{i\chi} M_{2}^{1}$,
where $\gamma$, $\delta$ and $\chi$ are such that $U_{\cal E}$ is
a unitary operation. If Alice and Bob choose $U_{\cal E}$ such
that all these requirements are not verified, then the probability
of successful tampering will be strictly less than one,
independently of Eve's TPCP map.

\subsubsection{Measurement}

Let us assume now that, instead of performing a predetermined
quantum operation on the message sent by Alice, Eve makes a
measurement on it trying to gain information about the key.  If
she were able to collapse the state of the key in a known
unentangled pure state, she could throw away Alice's message and
prepare and send to Bob an unauthentic new one that would pass his
test with certainty. Since Eve knows how the protocol works, she
would achieve this if she could distinguish perfectly between the
two terms on the right-hand side of (\ref{prot1}).

In order to avoid this attack, Alice and Bob must choose $U_{\cal E}$ such
that the set of states $\{|\phi_{i}\rangle, U_{\cal E}|\phi_{i}\rangle\}$,
with ${i=0,1}$, is not orthogonal. Owing to the orthogonality of
$|\phi_{0}\rangle$ and $|\phi_{1}\rangle$, this requirement can be
rewritten as $\langle\phi_{i}|U_{\cal E}|\phi_{j}\rangle\neq{}0$ for, at
least, one $i$ and $j$, with $i,j\in\{0,1\}$.  With the block notation
introduced in previous sections, this condition can be expressed as
$|M_{0}^{0}|>0$ or $|M_{0}^{1}|>0$. Although no secrecy is necessary
for secure authentication, note that if $\langle\phi_{i}|U_{\cal
E}|\phi_{j}\rangle\neq{}0$, with $i\neq{}j$, the quantum authentication
scheme also provides, in some sense, data encryption, since there is
a probability bigger than zero, of Eve not determining which message
Alice sent.

\subsection{Discussion}

MACs are used to detect any attempt to modify the transmitted data
by an undesired third party. In this section we have concentrated
on several types of attacks which, we believe, are the most
demanding.  We have shown that, in order to avoid the forgery
strategies studied, Alice and Bob should agree to choose $U_{\cal
E}$ such that the following conditions are satisfied:

\begin{enumerate}
\item If $|\bar{M}_{0}^{1}\bar{M}_{0}^{0\dag}|=0$, then
$|\bar{M}_{0}^{0}|^{2}<1$ and $|\bar{M}_{0}^{1}|^{2}<1$.

\item If $|\bar{M}_{0}^{1}\bar{M}_{0}^{0\dag}|\neq 0$, then equation
(\ref{EC_GORDA}) must be verified.

\item $M_{0}^{0}\neq{}e^{i\gamma}S(e^{i\delta})\sigma_xM_{0}^{1}$, or
$M_{2}^{0\dag}M_{2}^{1}\neq 0$ and $M_{2}^{0}\neq e^{i\chi} M_{2}^{1}$.

\item $|M_{0}^{0}|>0$ or $|M_{0}^{1}|>0$.
\end{enumerate}

Of these four conditions, it is straightforward to see, however,
that the last one, obtained in order to avoid the determination of
the key by measurement, is redundant, since the fulfillment of the
third condition leads to the fourth one.

After examining the three remaining conditions, two questions
arise: (i) Can a unitary operation simultaneously fulfill these
three restrictions? and, (ii) If the answer is yes, what is the
optimum $U_{\cal E}$?  Perhaps the easiest way to answer the first
question is with a trivial example. If, for instance,
$\bar{M}_{0}^{0}=(0.5\quad 0.5)$ and $\bar{M}_{0}^{1}=(0\quad 0)$,
it is straightforward to construct a unitary operation with its
first block equal to $M_{0}$. Moreover, it is evident that all the
above conditions are satisfied by this matrix. As for the second question,
it is an important open issue that we plan to address in the future.
First one should establish some appropriate criterion according to which
obtain such an optimum $U_\E$.  When we analyzed no-message attacks, we
showed that, selecting an appropriate $U_\E$, $P_f$ can be made $1/2$
regardless of Eve's strategy. Nevertheless, it is straightforward to
see that this particular unitary quantum operation makes $P_f^\prime$
one, thus making the protocol vulnerable. Therefore, it seems that the
optimization should result from a balance of the different forgery
strategies considered.

Finally, one interesting property of this class of quantum
authentication protocols is that it provides the possibility of
reusing the authentication keys: If there is no forgery, then after
Alice's encoding and Bob's decoding processes the state of the key
remains intact. Thus, if the authentication procedure is successful,
in principle Alice and Bob could retain the entangled key and reuse it
in the next run of the protocol. The presence of Eve, however, cannot be
despised. She could try to entangle an ancilla system with the quantum
authentication key generating a global state of the form:

\begin{equation}\label{KEY_FALSE}
|\phi\rangle_{ABE}=\alpha|01\rangle_{AB}|\phi\rangle_{E}-\beta|10\rangle_{AB}
|\phi_{\perp}\rangle_{E},
\end{equation}

\noindent with
$|\phi\rangle_{ABE}\in\mathcal{K}\otimes\mathcal{A}$, where
$\mathcal{K}$ and $\mathcal{A}$ denote the state spaces of the key
and the ancilla systems, respectively; $|\phi\rangle_{E}$ and
$|\phi_{\perp}\rangle_{E}$ represent two arbitrary orthonormal
states in $\mathcal{A}$; and $\alpha$ and $\beta$ are two
arbitrary complex numbers satisfying $|\alpha|^2+|\beta|^2=1$. If
equation (\ref{KEY_FALSE}) is verified, Eve could always forge
messages when the key is reused, just reproducing Alice's encoding
process, but employing her ancilla as the control of the quantum
operation.

If we assume that Eve has access only to the quantum channel
between Alice and Bob, which we believe is a reasonable
assumption, then Eve could try to obtain (\ref{KEY_FALSE}) in two
different ways.  She could prepare a quantum message and send it
to Bob, or she could manipulate the message sent by Alice. The
first possibility can be neglected, since, if $U_\E$ satisfies the
conditions enumerated above, Eve cannot know when a run of the
protocol has been successful. As for the second possibility, it
must not be confused with the one previously analyzed when dealing
with TPCP maps. Now Eve does not need to convert
$|\phi_{0}\rangle$ into $|\phi_{1}\rangle$ and vice versa. She can
prepare $|\psi\rangle_{E}\in\mathcal{A}$ and apply a unitary
operation $U_{\mathcal{E}\otimes\mathcal{A}}$ of the form:

\begin{equation}
U_{\mathcal{E}\otimes\mathcal{A}}
\left[
\frac{1}{\sqrt{2}}
(|\phi_{i}\rangle|01\rangle_{AB}-U_{\cal E}|\phi_{i}\rangle|10\rangle_{AB})
\otimes{}|\psi\rangle_{E}
\right],
\end{equation}

\noindent trying to achieve $U_{\mathcal{E}\otimes\mathcal{A}}
(|\phi_i\rangle|\psi\rangle_{E})=(\alpha|\phi_{i}\rangle+
\beta|\phi_{j}\rangle)|\phi\rangle_{E}$
and $U_{\mathcal{E}\otimes\mathcal{A}} (U_{\cal
E}|\phi_i\rangle|\psi\rangle_{E})=(\gamma{}U_{\cal E}|\phi_{i}\rangle+
\delta{}U_{\cal E}|\phi_{j}\rangle)|\phi_{\perp}\rangle_{E}$, with
$i,j\in\{0,1\}$, and $\alpha,\beta,\gamma,\delta$ some complex parameters
such that $|\alpha|^2+|\beta|^2=|\gamma|^2+|\delta|^2=1$. If $U_{\cal E}$
is chosen such that $\langle \phi_i|U_\E|\phi_i\rangle \neq 0$, for some
$i \in \{0,1\}$, then $|\phi_{i}\rangle$ and $U_{\cal E}|\phi_{i}\rangle$
are not orthogonal for at least one value of $i$. Therefore, and since
the inner product of states is preserved by any unitary operation,
these conditions are impossible to fulfill. This means that equation
(\ref{KEY_FALSE}) cannot be achieved with certainty.

Nevertheless, and although key recycling is in principle possible,
it should be noticed that the security of the authentication
protocols presented may be drastically reduced. As suggested in
\cite{LEUNG_2000}, security does not depend on the use of
entanglement, but on the possibility of detecting Eve's presence in
the quantum ciphertext. As we have seen, these authentication
schemes can detect Eve with a certain probability, but there is
also a chance that Eve remains undetected.

\section{Conclusion}
\label{CONCLUSIONES}

We have presented a broad class of quantum authentication
protocols that, making use of just one qubit as the authentication
key, allow the authentication of binary classical messages with a
probability of successful forgery less than one. All parties,
including the forger, may have full knowledge about all aspects of
the protocol; however, it requires sharing a previous secret (in
the form of an entangled pair of particles, or a classical bit),
and an ideal quantum channel between the partners.

We have described several types of possible attacks and shown that
careful selection of the quantum transformation performed by the
communicating parties makes the protocol secure against these
attacks. However, a further more extensive security analysis in a
more realistic scenario (a non-perfect channel), as well as the
derivation of the optimum $U_\E$ in such circumstances, is
needed.

Finally, we have also shown that the protocol authentication keys can
be reused.  However, this reduces the security of the protocol.

\section*{Acknowledgments}

The authors acknowledge Howard Barnum for sharing with them his work on
authentication, and Esther P\'erez, Ad\'an Cabello and Debbie Leung for
their insightful comments. This work was partially supported by Xunta
de Galicia, Spain (grant n.\ PGIDT00PXI322060PR).


\begin{thebibliography}{10}
\expandafter\ifx\csname bibnamefont\endcsname\relax
  \def\bibnamefont#1{#1}\fi
\expandafter\ifx\csname bibfnamefont\endcsname\relax
  \def\bibfnamefont#1{#1}\fi
\expandafter\ifx\csname url\endcsname\relax
  \def\url#1{\texttt{#1}}\fi
\expandafter\ifx\csname urlprefix\endcsname\relax\def\urlprefix{URL }\fi
\providecommand{\bibinfo}[2]{#2}
\providecommand{\eprint}[2][]{\url{#2}}

\bibitem{DENNING_1983}
\bibinfo{author}{\bibfnamefont{D.~E.~R.} \bibnamefont{Denning}},
  \emph{\bibinfo{title}{Cryptography and Data Security}}
  (\bibinfo{publisher}{Addison-Wesley Publishing Company},
  \bibinfo{year}{1983}).

\bibitem{SCHNEIER_1996}
\bibinfo{author}{\bibfnamefont{B.}~\bibnamefont{Schneier}},
  \emph{\bibinfo{title}{Applied Cryptography}} (\bibinfo{publisher}{John Wiley
  \& Sons, Inc.}, \bibinfo{year}{1996}), \bibinfo{edition}{2nd} ed.

\bibitem{GRUSKA_1999}
\bibinfo{author}{\bibfnamefont{J.}~\bibnamefont{Gruska}},
  \emph{\bibinfo{title}{Quantum Computing}} (\bibinfo{publisher}{McGraw-Hill},
  \bibinfo{year}{1999}).

\bibitem{LO_1999_a}
\bibinfo{author}{\bibfnamefont{H.}~\bibnamefont{Lo}},
  \bibinfo{author}{\bibfnamefont{S.}~\bibnamefont{Popescu}}, \bibnamefont{and}
  \bibinfo{author}{\bibfnamefont{T.}~\bibnamefont{Spiller}},
  \emph{\bibinfo{title}{Introduction to quantum computation and information}}
  (\bibinfo{publisher}{World Scientific}, \bibinfo{address}{Singapore},
  \bibinfo{year}{1999}).

\bibitem{NIELSEN_2000}
\bibinfo{author}{\bibfnamefont{M.~A.} \bibnamefont{Nielsen}} \bibnamefont{and}
  \bibinfo{author}{\bibfnamefont{I.~L.} \bibnamefont{Chuang}},
  \emph{\bibinfo{title}{Quantum Computation and Quantum Information}}
  (\bibinfo{publisher}{Cambridge University Press},
  \bibinfo{address}{Cambridge}, \bibinfo{year}{2000}).

\bibitem{WIESNER_1983}
\bibinfo{author}{\bibfnamefont{S.~J.} \bibnamefont{Wiesner}},
  \bibinfo{journal}{SIGACT News} \textbf{\bibinfo{volume}{15}},
  \bibinfo{pages}{78} (\bibinfo{year}{1983}), \bibinfo{note}{original
  manuscript written ca. 1970}.

\bibitem{BENNETT_1982}
\bibinfo{author}{\bibfnamefont{C.~H.} \bibnamefont{Bennett}},
  \bibinfo{author}{\bibfnamefont{G.}~\bibnamefont{Brassard}},
  \bibinfo{author}{\bibfnamefont{S.}~\bibnamefont{Breidbart}},
  \bibnamefont{and} \bibinfo{author}{\bibfnamefont{S.}~\bibnamefont{Wiesner}},
  in \emph{\bibinfo{booktitle}{Advances in Cryptology: Proceedings of Crypto
  82}} (\bibinfo{publisher}{Plenum Press}, \bibinfo{address}{New York},
  \bibinfo{year}{1982}), pp. \bibinfo{pages}{267--275}.

\bibitem{BENNETT_1992}
\bibinfo{author}{\bibfnamefont{C.~H.} \bibnamefont{Bennett}} \bibnamefont{and}
  \bibinfo{author}{\bibfnamefont{J.}~\bibnamefont{Smolin}},
  \bibinfo{journal}{Journal of Cryptology}
  \textbf{\bibinfo{volume}{5}}(\bibinfo{number}{1}), \bibinfo{pages}{3}
  (\bibinfo{year}{1992}).

\bibitem{MULLER_1993}
\bibinfo{author}{\bibfnamefont{A.}~\bibnamefont{Muller}},
  \bibinfo{author}{\bibfnamefont{J.}~\bibnamefont{Breguet}}, \bibnamefont{and}
  \bibinfo{author}{\bibfnamefont{N.}~\bibnamefont{Gisin}},
  \bibinfo{journal}{Europhysics Letters}
  \textbf{\bibinfo{volume}{23}}(\bibinfo{number}{6}), \bibinfo{pages}{383}
  (\bibinfo{year}{1993}).

\bibitem{MARAND_1995}
\bibinfo{author}{\bibfnamefont{C.}~\bibnamefont{Marand}} \bibnamefont{and}
  \bibinfo{author}{\bibfnamefont{P.~D.} \bibnamefont{Townsend}},
  \bibinfo{journal}{Optics Letters}
  \textbf{\bibinfo{volume}{20}}(\bibinfo{number}{16}), \bibinfo{pages}{1695}
  (\bibinfo{year}{1995}).

\bibitem{HUGHES_1995}
\bibinfo{author}{\bibfnamefont{R.~J.} \bibnamefont{Hughes}},
  \bibinfo{author}{\bibfnamefont{D.~M.} \bibnamefont{Alde}},
  \bibinfo{author}{\bibfnamefont{P.}~\bibnamefont{Dyer}},
  \bibinfo{author}{\bibfnamefont{G.~G.} \bibnamefont{Luther}},
  \bibinfo{author}{\bibfnamefont{G.~L.} \bibnamefont{Morgan}},
  \bibnamefont{and} \bibinfo{author}{\bibfnamefont{M.}~\bibnamefont{Schauer}},
  \bibinfo{journal}{Contemporary Physics}
  \textbf{\bibinfo{volume}{36}}(\bibinfo{number}{3}), \bibinfo{pages}{149}
  (\bibinfo{year}{1995}).

\bibitem{BUTTLER_2000}
\bibinfo{author}{\bibfnamefont{W.~T.} \bibnamefont{Buttler}},
  \bibinfo{author}{\bibfnamefont{R.~J.} \bibnamefont{Hughes}},
  \bibinfo{author}{\bibfnamefont{S.~K.} \bibnamefont{Lamoreaux}},
  \bibinfo{author}{\bibfnamefont{G.~L.} \bibnamefont{Morgan}},
  \bibinfo{author}{\bibfnamefont{J.~E.} \bibnamefont{Nordholt}},
  \bibnamefont{and} \bibinfo{author}{\bibfnamefont{C.~G.}
  \bibnamefont{Peterson}}, \bibinfo{journal}{Physical Review Letters}
  \textbf{\bibinfo{volume}{84}}, \bibinfo{pages}{5652} (\bibinfo{year}{2000}).

\bibitem{TITTEL_2000}
\bibinfo{author}{\bibfnamefont{W.}~\bibnamefont{Tittel}},
  \bibinfo{author}{\bibfnamefont{J.}~\bibnamefont{Brendel}},
  \bibinfo{author}{\bibfnamefont{H.}~\bibnamefont{Zbinden}}, \bibnamefont{and}
  \bibinfo{author}{\bibfnamefont{N.}~\bibnamefont{Gisin}},
  \bibinfo{journal}{Physical Review Letters} \textbf{\bibinfo{volume}{84}},
  \bibinfo{pages}{4737} (\bibinfo{year}{2000}).

\bibitem{NAIK_2000}
\bibinfo{author}{\bibfnamefont{D.~S.} \bibnamefont{Naik}},
  \bibinfo{author}{\bibfnamefont{C.~G.} \bibnamefont{Peterson}},
  \bibinfo{author}{\bibfnamefont{A.~G.} \bibnamefont{White}},
  \bibinfo{author}{\bibfnamefont{A.~J.} \bibnamefont{Berglund}},
  \bibnamefont{and} \bibinfo{author}{\bibfnamefont{P.~G.} \bibnamefont{Kwiat}},
  \bibinfo{journal}{Physical Review Letters} \textbf{\bibinfo{volume}{84}},
  \bibinfo{pages}{4733} (\bibinfo{year}{2000}).

\bibitem{JENNEWEIN_2000}
\bibinfo{author}{\bibfnamefont{T.}~\bibnamefont{Jennewein}},
  \bibinfo{author}{\bibfnamefont{C.}~\bibnamefont{Simon}},
  \bibinfo{author}{\bibfnamefont{G.}~\bibnamefont{Weihs}},
  \bibinfo{author}{\bibfnamefont{H.}~\bibnamefont{Weinfurter}},
  \bibnamefont{and}
  \bibinfo{author}{\bibfnamefont{A.}~\bibnamefont{Zeilinger}},
  \bibinfo{journal}{Physical Review Letters} \textbf{\bibinfo{volume}{84}},
  \bibinfo{pages}{4729} (\bibinfo{year}{2000}).

\bibitem{BENNETT_1984}
\bibinfo{author}{\bibfnamefont{C.~H.} \bibnamefont{Bennett}} \bibnamefont{and}
  \bibinfo{author}{\bibfnamefont{G.}~\bibnamefont{Brassard}}, in
  \emph{\bibinfo{booktitle}{Proceedings of IEEE International Conference on
  Computers, Systems and Signal Processing}} (\bibinfo{publisher}{IEEE Press},
  \bibinfo{address}{New York}, \bibinfo{year}{1984}), pp.
  \bibinfo{pages}{175--179}.

\bibitem{EKERT_1991}
\bibinfo{author}{\bibfnamefont{A.~K.} \bibnamefont{Ekert}},
  \bibinfo{journal}{Physical Review Letters}
  \textbf{\bibinfo{volume}{67}}(\bibinfo{number}{6}), \bibinfo{pages}{661}
  (\bibinfo{year}{1991}).

\bibitem{BENNETT_1992_a}
\bibinfo{author}{\bibfnamefont{C.~H.} \bibnamefont{Bennett}},
  \bibinfo{journal}{Physical Review Letters}
  \textbf{\bibinfo{volume}{68}}(\bibinfo{number}{21}), \bibinfo{pages}{3121}
  (\bibinfo{year}{1992}).

\bibitem{LJUNGGREN_2000}
\bibinfo{author}{\bibfnamefont{D.}~\bibnamefont{Ljunggren}},
  \bibinfo{author}{\bibfnamefont{M.}~\bibnamefont{Bourennane}},
  \bibnamefont{and} \bibinfo{author}{\bibfnamefont{A.}~\bibnamefont{Karlsson}},
  \bibinfo{journal}{Physical Review~A} \textbf{\bibinfo{volume}{62}},
  \bibinfo{pages}{022305} (\bibinfo{year}{2000}).

\bibitem{ZENG_2000}
\bibinfo{author}{\bibfnamefont{G.}~\bibnamefont{Zeng}} \bibnamefont{and}
  \bibinfo{author}{\bibfnamefont{W.}~\bibnamefont{Zhang}},
  \bibinfo{journal}{Physical Review~A} \textbf{\bibinfo{volume}{61}},
  \bibinfo{pages}{022303} (\bibinfo{year}{2000}).

\bibitem{CREPEAU_1995}
\bibinfo{author}{\bibfnamefont{C.}~\bibnamefont{Cr\'epeau}} \bibnamefont{and}
  \bibinfo{author}{\bibfnamefont{L.}~\bibnamefont{Salvail}}, in
  \emph{\bibinfo{booktitle}{Advances in Cryptology: Proceedings of Eurocrypt
  '95}} (\bibinfo{publisher}{Springer-Verlag}, \bibinfo{address}{Berlin},
  \bibinfo{year}{1995}), pp. \bibinfo{pages}{133--146}.

\bibitem{DUSEK_1999}
\bibinfo{author}{\bibfnamefont{M.}~\bibnamefont{Dusek}},
  \bibinfo{author}{\bibfnamefont{O.}~\bibnamefont{Haderka}},
  \bibinfo{author}{\bibfnamefont{M.}~\bibnamefont{Hendrych}}, \bibnamefont{and}
  \bibinfo{author}{\bibfnamefont{R.}~\bibnamefont{Myska}},
  \bibinfo{journal}{Physical Review~A}
  \textbf{\bibinfo{volume}{60}}(\bibinfo{number}{1}), \bibinfo{pages}{149}
  (\bibinfo{year}{1999}).

\bibitem{barnum_99}
\bibinfo{author}{\bibfnamefont{H.}~\bibnamefont{Barnum}},
  \emph{\bibinfo{title}{Quantum secure identification using entanglement and
  catalysis}}, \bibinfo{note}{quant-ph/9910072}.

\bibitem{ZENG_2000_a}
\bibinfo{author}{\bibfnamefont{G.}~\bibnamefont{Zeng}} \bibnamefont{and}
  \bibinfo{author}{\bibfnamefont{G.}~\bibnamefont{Guo}},
  \emph{\bibinfo{title}{Quantum authentication protocol}},
  \bibinfo{note}{quant-ph/0001046}.

\bibitem{JENSEN_2000}
\bibinfo{author}{\bibfnamefont{J.~G.} \bibnamefont{Jensen}} \bibnamefont{and}
  \bibinfo{author}{\bibfnamefont{R.}~\bibnamefont{Schack}},
  \emph{\bibinfo{title}{Quantum authentication and key distribution using
  catalysis}}, \bibinfo{note}{quant-ph/0003104}.

\bibitem{zhang_00}
\bibinfo{author}{\bibfnamefont{Y.-S.} \bibnamefont{Zhang}},
  \bibinfo{author}{\bibfnamefont{C.-F.} \bibnamefont{Li}}, \bibnamefont{and}
  \bibinfo{author}{\bibfnamefont{G.-C.} \bibnamefont{Guo}},
  \emph{\bibinfo{title}{Quantum authentication using entangled state}},
  \bibinfo{note}{quant-ph/0008044}.

\bibitem{CARTER_1979}
\bibinfo{author}{\bibfnamefont{J.~L.} \bibnamefont{Carter}} \bibnamefont{and}
  \bibinfo{author}{\bibfnamefont{M.~N.} \bibnamefont{Wegman}},
  \bibinfo{journal}{J. Computer and System Sciences}
  \textbf{\bibinfo{volume}{18}}, \bibinfo{pages}{143} (\bibinfo{year}{1979}).

\bibitem{WEGMAN_1981}
\bibinfo{author}{\bibfnamefont{M.~N.} \bibnamefont{Wegman}} \bibnamefont{and}
  \bibinfo{author}{\bibfnamefont{J.~L.} \bibnamefont{Carter}},
  \bibinfo{journal}{J. Computer and System Sciences}
  \textbf{\bibinfo{volume}{22}}, \bibinfo{pages}{265} (\bibinfo{year}{1981}).

\bibitem{barnum_2001}
\bibinfo{author}{\bibfnamefont{H.}~\bibnamefont{Barnum}},
  \emph{\bibinfo{title}{Quantum message authentication codes}},
  \bibinfo{note}{quant-ph/0103123}.

\bibitem{HORN_1985}
\bibinfo{author}{\bibfnamefont{R.~A.} \bibnamefont{Horn}} \bibnamefont{and}
  \bibinfo{author}{\bibfnamefont{C.~R.} \bibnamefont{Johnson}},
  \emph{\bibinfo{title}{Matrix Analysis}} (\bibinfo{publisher}{Cambridge
  University Press}, \bibinfo{address}{Cambridge}, \bibinfo{year}{1985}).

\bibitem{LEUNG_2000}
\bibinfo{author}{\bibfnamefont{D.~W.} \bibnamefont{Leung}},
  \emph{\bibinfo{title}{Quantum {V}ernam cipher}},
  \bibinfo{note}{quant-ph/0012077}.

\end{thebibliography}

\end{document}